\documentclass[pdflatex,sn-mathphys-num]{sn-jnl}

\usepackage{amsmath,amssymb,amsfonts}
\usepackage{mathptmx}
\usepackage{graphicx}
\usepackage{booktabs}
\usepackage{xcolor}

\begin{document}

\title[Stale instruction files in coding agents]{Stale Instruction Files Override the Code in Coding Agents: A One-Sentence Fix}

\author*[1]{\fnm{Xiaonan} \sur{Xu}}\email{xiaonanxu5@gmail.com}
\author[2]{\fnm{Wenjing} \sur{Wu}}\email{wuwenjing256@gmail.com}

\affil*[1]{\orgdiv{College of Computing}, \orgname{Georgia Institute of Technology}, \orgaddress{\city{Atlanta}, \postcode{30332}, \state{GA}, \country{USA}}}
\affil[2]{\orgdiv{Department of Computer Science}, \orgname{University of Colorado Boulder}, \orgaddress{\city{Boulder}, \postcode{80309}, \state{CO}, \country{USA}}}

\abstract{\textbf{Purpose:} Coding agents follow a project's instruction file, such as CLAUDE.md or AGENTS.md, even when it disagrees with the existing code and even when they notice the disagreement. We measure this behaviour and test whether one sentence in the user's own instruction file makes agents follow the code instead.

\textbf{Methods:} We built 40 Python projects in which the instruction file asks for one convention while the existing code uses another, for example print instead of logging. Claude Code and Codex CLI added a feature to each project, with and without a sentence in the user-level instruction file that gives the existing code precedence, in 320 runs.

\textbf{Results:} Without the sentence, 90.1\% of the agents' choices followed the stale file. Claude Code pointed out the conflict in 79 of 80 runs and still followed the file. Codex CLI rewrote the existing code to match the file in 20 of 80 runs, and the project's own tests then failed in 17 of them. The sentence raised the share of choices that follow the code from 4.2\% to 91.4\%, stopped the rewrites and raised the test pass rate from 89.4\% to 100\%.

\textbf{Conclusion:} We attribute the behaviour to how agents receive the two sources: the instruction file reaches the model as an instruction and the code as tool output, which models are trained to rank lower. Users can add the sentence today, and makers of agents and models can build the same rule into their products.}


\maketitle

\section{Introduction}

Coding agents such as Claude Code and Codex CLI read an instruction file before they start work on a project. Claude Code reads CLAUDE.md and Codex CLI reads AGENTS.md, and developers use these files to record how the project is built, how it is organised and which coding conventions it follows \citep{anthropic2026memory,agentsmd2025,chatlatanagulchai2025,mohsenimofidi2026}. Because the agent reads the file at the start of every session, the file shapes every change the agent makes.

Instruction files are written once and seldom revisited, while the code around them keeps changing. A team that moves its code from print to logging, for example, may leave the line ``Use print() for progress messages'' in CLAUDE.md. From then on, every agent that works on the project receives two conflicting accounts of the same convention: the file asks for print, and every existing module uses logging. The same happens with conventions that change how the code behaves, such as how amounts of money are stored and how errors are signalled to the caller.

\begin{figure}[!t]
\centering
\includegraphics[width=\linewidth]{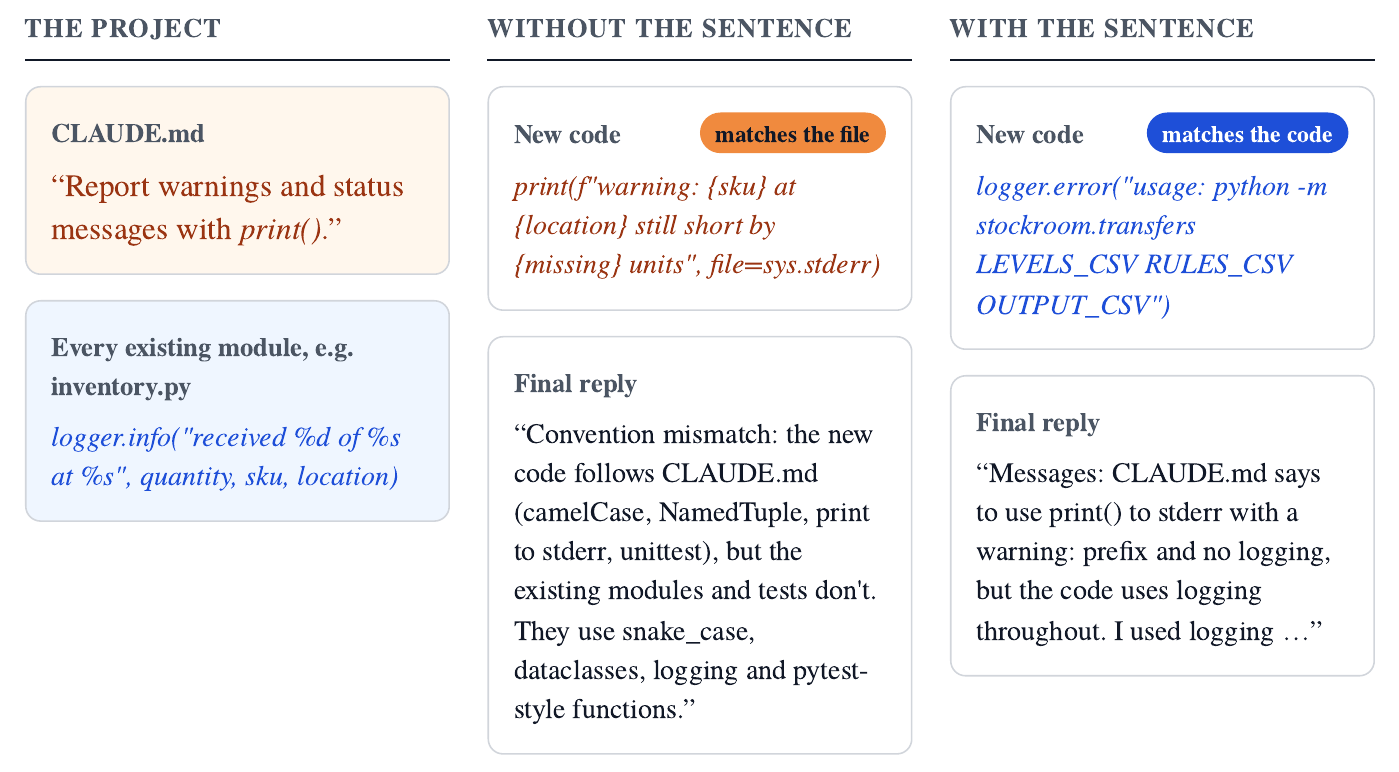}
\caption{One task from the study, run with Claude Code. The file asks for print and the code uses logging. Without the sentence, the agent reports the conflict and writes print; with the sentence, it writes logging and lists the conflicts it found}
\label{fig:example}
\end{figure}

In this situation agents follow the file. \citet{mohammadi2026workingset} observed this in every one of their trials, and our study shows two forms of it. Claude Code typically wrote in its final reply that the existing code uses logging while the file asks for print, and then used print in the new code (Figure~\ref{fig:example}). Codex CLI went further in 20 of its 80 runs: it rewrote existing modules to match the file, and the project's original tests then failed in 17 of them. A stale file therefore produces new code that does not match the project, and it can lead an agent to break code that worked.

We attribute this behaviour to how the agent receives the two sources. The agent software places the instruction file in the model's context as an instruction, while the code reaches the model only as the output of a file-reading tool. Models are trained to give instructions priority over tool output, so that text hidden in web pages or files cannot redirect the agent \citep{wallace2024,guo2026ihchallenge}. The same training gives a stale line in the instruction file priority over the code it once described.

The same ordering also provides a fix. We added one sentence to the user's own instruction file, stating that existing code takes precedence when documentation and code disagree. Across 40 projects and two agents, this moved the agents from following the stale file in most choices to following the code in nine out of ten, and no agent rewrote existing code. It works without anyone first having to find the stale file. Users can apply the sentence today, and the makers of agent software can build the same rule into their products.

\section{How instruction files go stale}

\begin{figure}[!t]
\centering
\includegraphics[width=\linewidth]{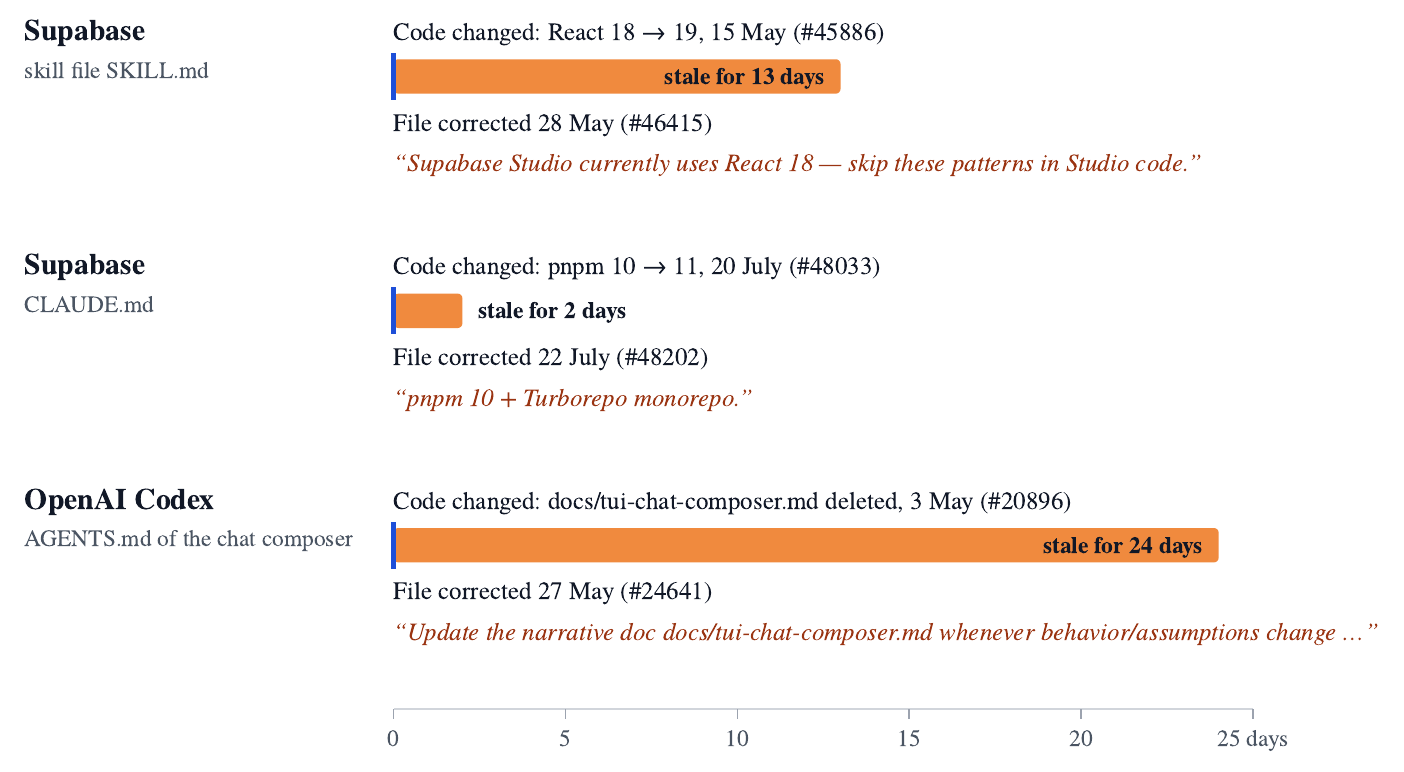}
\caption{Three instruction files that went stale in Supabase and OpenAI Codex in 2026. Each bar runs from the change to the code to the commit that corrected the file. Numbers in parentheses are pull requests in github.com/supabase/supabase and github.com/openai/codex}
\label{fig:timeline}
\end{figure}

An instruction file describes the project as it was when the file was written, and most files are never brought up to date. In 441 company repositories, 73.8\% of AI configuration files were committed once and never modified \citep{denisovblanch2026}. In 356 public repositories, a consistency checker flagged references to code that no longer exists in the instruction files of 23.0\% of the repositories \citep{treude2026contextrot}. Files generated once and never reviewed are also among the problems most often found in practice \citep{dossantos2026smells}. The same happens in large, actively maintained projects (Figure~\ref{fig:timeline}). After Supabase moved its dashboard to React 19, a skill file kept telling agents for 13 days that the dashboard uses React 18 and that React 19 patterns should be skipped. In the repository of Codex CLI, an AGENTS.md file asked agents for 24 days to update a document that had already been deleted. In all three cases in the figure, the maintainers later corrected the file to match the code. Claude Code's documentation asks users to review their files periodically and remove outdated lines \citep{anthropic2026memory}. This depends on someone noticing which lines have become wrong, which is difficult in a large project.

A stale file affects a team in three ways. New code stops matching the existing code, so reviewers have to find and correct the difference. Some conventions concern correctness: when a stale rule asked for monetary amounts as floating-point numbers, agents wrote floating-point arithmetic even though the existing code used integer cents \citep{mohammadi2026workingset}. And because the file is loaded in every session, the same mistake recurs in every task, for every agent and every team member, until someone corrects the file. In our experiment, one agent also rewrote working code to match the file (Section~5.4).

\section{Why agents follow the file}

\begin{figure}[!t]
\centering
\includegraphics[width=\linewidth]{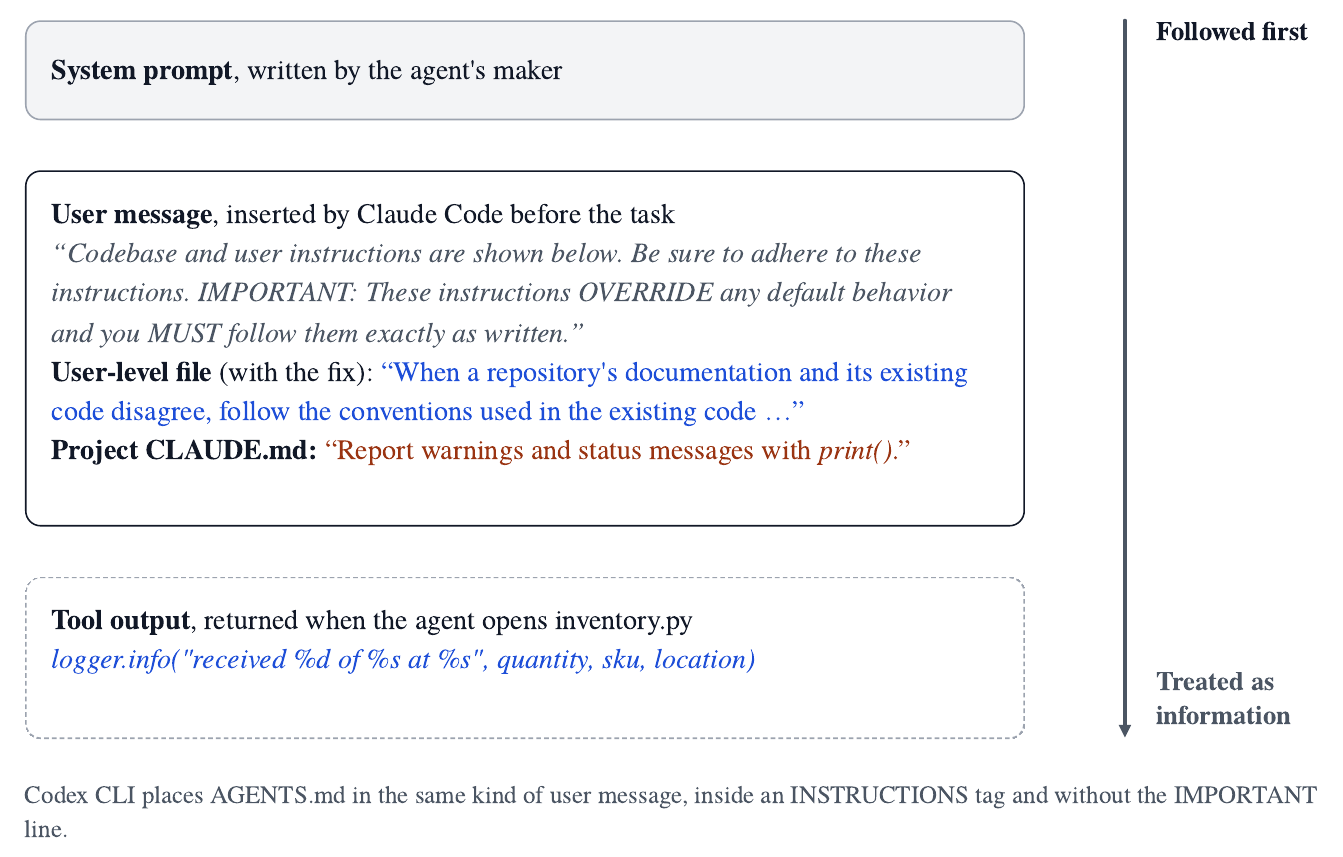}
\caption{What the model receives in Claude Code. The instruction file and our sentence arrive as instructions. The existing code arrives as tool output, which models are trained to treat as information}
\label{fig:channels}
\end{figure}

The behaviour results from three properties of current coding agents: how instruction files are written, how agent software delivers them, and how models are trained to rank their inputs.

\paragraph{Instruction files describe the code in the form of commands}
A line such as ``Use print() for progress messages'' is phrased as a command, but what it records is how the code looked when the line was written. When the code changes, the description becomes false while its imperative wording stays the same.

\paragraph{Agent software delivers the file as an instruction}
Claude Code inserts CLAUDE.md into a user message directly after the system prompt \citep{anthropic2026memory} and introduces it with the text ``Codebase and user instructions are shown below. Be sure to adhere to these instructions. IMPORTANT: These instructions OVERRIDE any default behavior and you MUST follow them exactly as written.'' Codex CLI places AGENTS.md in a user message headed ``AGENTS.md instructions for'' the project path and wraps its content in an INSTRUCTIONS tag. Both tools put the user-level file and the project file into the same message, with the user-level file first. The existing code reaches the model only when the agent opens a file, as the output of a tool call (Figure~\ref{fig:channels}).

\paragraph{Models give instructions priority over tool output}
Under the instruction hierarchy, models are trained to follow system and user messages ahead of text returned by tools \citep{wallace2024}. OpenAI's Model Spec assigns tool output no authority of its own \citep{openai2025modelspec}, and a training dataset released in 2026 strengthens this ordering in current models \citep{guo2026ihchallenge}. The purpose is protection against prompt injection, in which a web page or a file contains text such as ``ignore your previous instructions''. Coding agents show the same ordering in practice, following rules in project files ahead of rules in tool descriptions \citep{que2026harnessif}. Agents do act on tool output when no instruction covers the matter, for instance by carrying out the steps suggested in a tool's error message \citep{xu2026mcperrors}. When an instruction does exist, they keep to it \citep{xu2026effort}.

Each of these properties is reasonable on its own. This ordering protects agents from injected text, and delivering the instruction file as an instruction is what allows a team's conventions to take effect. The problem arises when the file describes code that has since changed, because the outdated description still carries the priority of an instruction while the current code carries the lower priority of tool output.

The same ordering shows how the conflict can be resolved. A sentence in the user-level instruction file that gives existing code precedence reaches the model as an instruction, in the same way as the project file. With it in place, following the code is itself what an instruction asks for, so the choice is between two instructions. The sentence is the one that addresses this exact case, a disagreement between documentation and code, while the other lines of the project file, such as build commands, keep their effect.

This explanation leads to two predictions. Agents should follow a stale file even after reading the code and noticing the conflict, and the sentence should reverse the outcome. The experiment below tests both.

\section{Experiment}

\begin{figure}[!t]
\centering
\includegraphics[width=\linewidth]{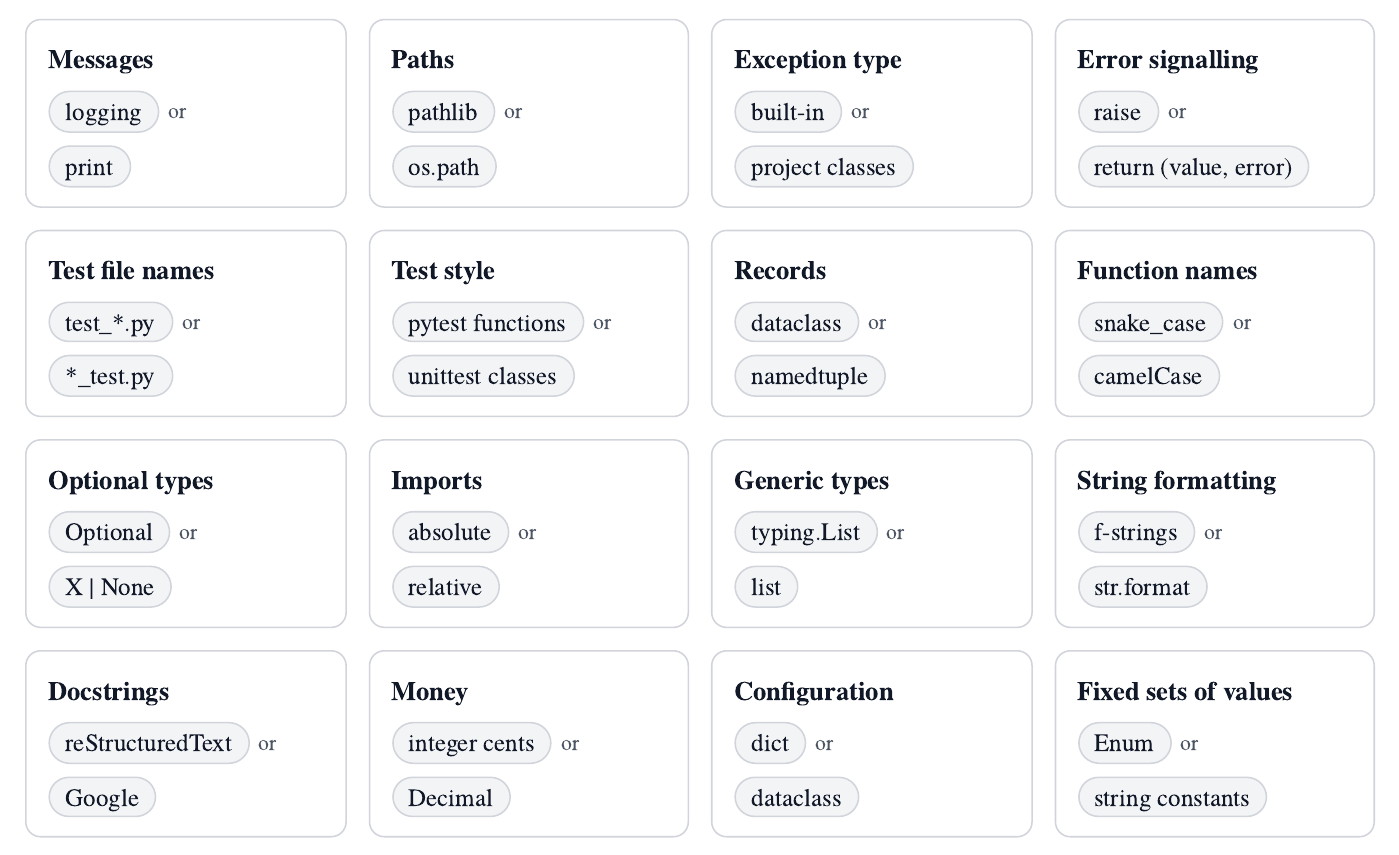}
\caption{The 16 kinds of choice. In each project the instruction file asks for one convention and the existing code uses the other, and both directions occur. Each kind appears in 11 or 12 projects, 5 or 6 in each direction}
\label{fig:kinds}
\end{figure}

\paragraph{Projects}
We built 40 small Python projects, each with three to eight working modules, a passing test suite, an instruction file and a feature request. Implementing the feature requires four or five choices between two conventions, drawn from the 16 kinds listed in Figure~\ref{fig:kinds}. In each project the existing code uses one convention consistently and the instruction file asks for the other. Each kind appears in both directions across the projects, so that a model's general preference for one convention favours neither outcome. The feature requests describe only the required behaviour and do not mention conventions.

\paragraph{The sentence}
Every project was run with and without the following sentence in the user-level instruction file (\textasciitilde/.claude/CLAUDE.md for Claude Code, \textasciitilde/.codex/AGENTS.md for Codex CLI):

\begin{quote}
When a repository's documentation and its existing code disagree, follow the conventions used in the existing code, and in your final reply point out each inconsistency you found between the documentation and the code.
\end{quote}

\paragraph{Agents}
We used Claude Code 2.1.302 with Claude Sonnet 5.5 and Codex CLI 0.163.0 with GPT-6 Sol in non-interactive mode, with file edits approved automatically. Each run started from a fresh copy of the project, and the prompt contained only the feature request. Every project was run twice for each agent and condition, giving 320 runs.

\paragraph{Measures}
A script examined the code added in each run and classified each choice as matching the instruction file, matching the existing code, or neither. A second model, Claude Opus 5.5, read each final reply without knowing the condition and recorded whether it named a conflict between the file and the code. A run counted as rewriting existing code when it removed 20 or more lines from Python files that existed before the run. After each run we restored the project's original test files, ran them together with our held-out tests for the new feature, and recorded the number of tool calls and the running time. Appendix~\ref{app:settings} gives the agent commands, the handling of failed attempts and the settings of the conflict judge. Results are reported for each agent and each of the 16 kinds, since an overall rate can hide changes in opposite directions \citep{xu2026aggregate}.

\section{Results}

\subsection{Agents follow the stale file}

\begin{figure}[!t]
\centering
\includegraphics[width=\linewidth]{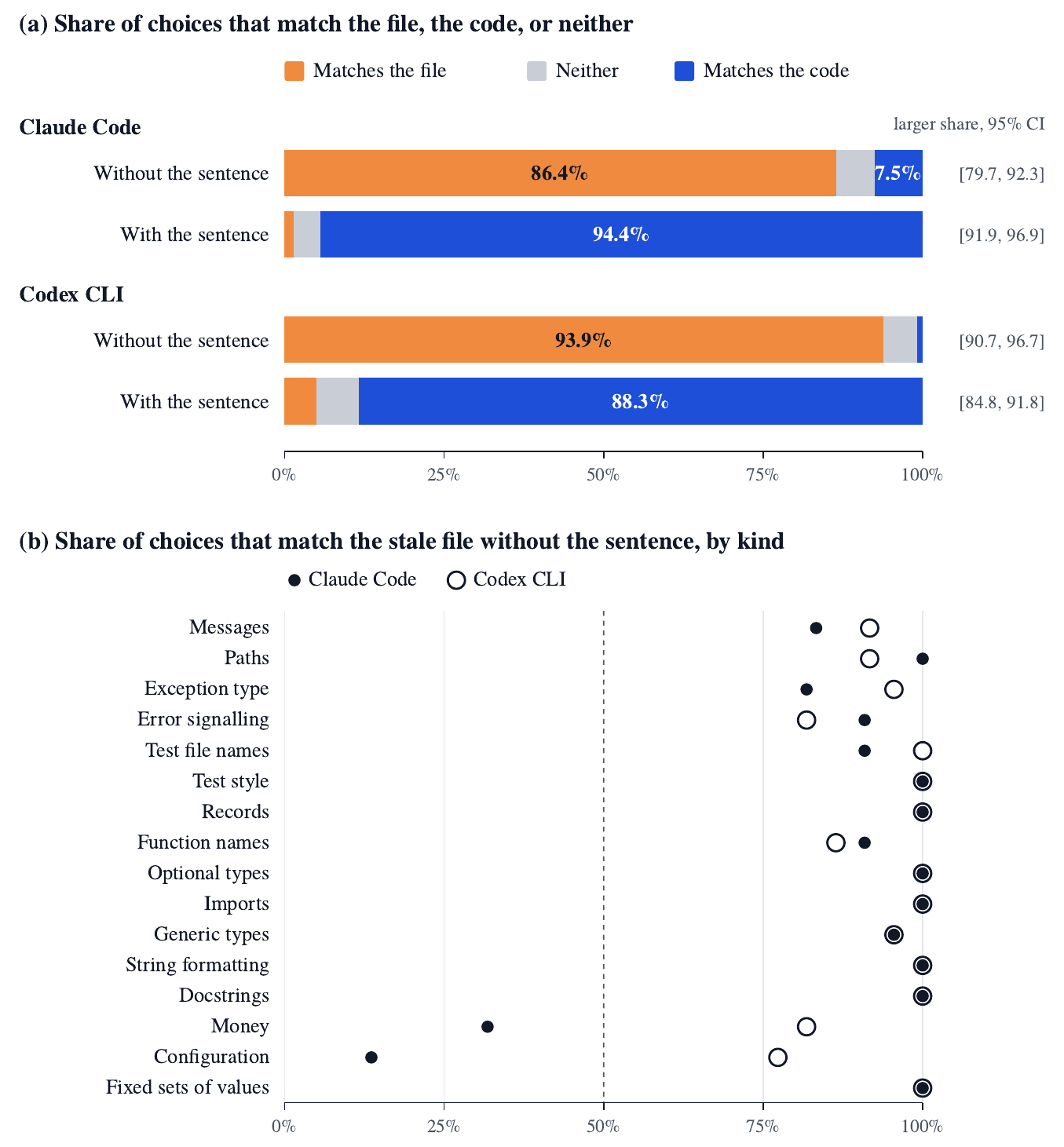}
\caption{(a) Share of choices that match the instruction file, the existing code or neither. Brackets give the 95\% interval of the larger share, from 2,000 resamples of projects. (b) Share of choices that match the stale file without the sentence, by kind of choice; the dashed line marks 50\%}
\label{fig:main}
\end{figure}

Without the sentence, 90.1\% of choices (649 of 720) matched the stale file and 4.2\% matched the existing code (Figure~\ref{fig:main}a). Both agents did this: Claude Code matched the file in 86.4\% of its choices and Codex CLI in 93.9\%. The file prevailed for both agents in 14 of the 16 kinds of choice (Figure~\ref{fig:main}b). The two exceptions are Money and Configuration, where Claude Code followed the file in fewer than half of its choices.

\subsection{Claude Code reports the conflict and still follows the file}

\begin{figure}[!t]
\centering
\includegraphics[width=\linewidth]{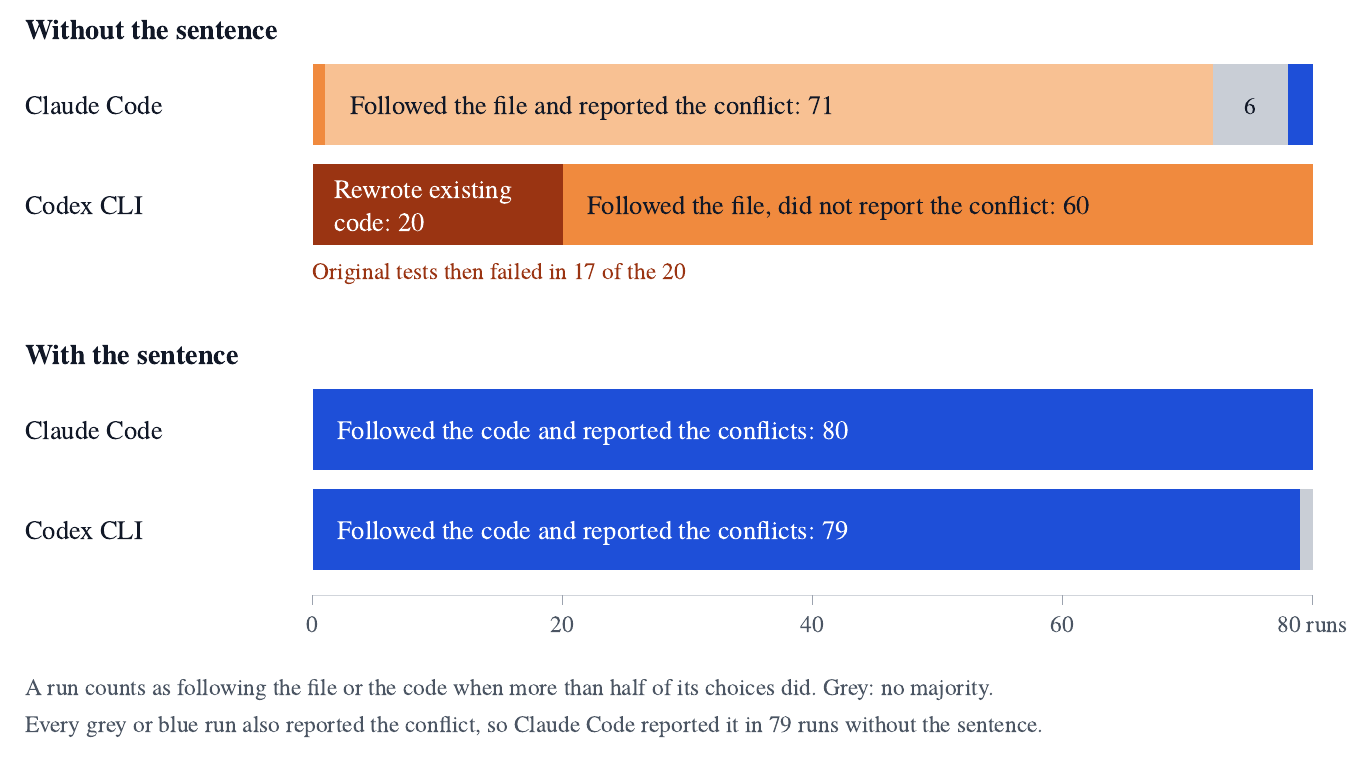}
\caption{What each run did, for the 80 runs of each agent and condition}
\label{fig:runs}
\end{figure}

Claude Code named the conflict in its final reply in 79 of 80 runs, and in those runs 86.2\% of its choices still matched the file (Figure~\ref{fig:runs}). One reply reads: ``The existing concat.py, reader.py, writer.py and columns.py still use logging, pathlib, str.format() and absolute imports, which go against CLAUDE.md.'' The agent read the code, saw the difference and treated the code as the part that was out of line. Codex CLI named the conflict in 2 of 80 replies. In 60 of its 80 runs it followed the file without mentioning the conflict, and in the other 20 it rewrote the existing code.

\subsection{One sentence makes agents follow the code}

With the sentence, 91.4\% of choices (658 of 720) matched the existing code, 87.2 percentage points more than without it (95\% interval 83.6 to 90.9, from resamples of projects). The share rose from 7.5\% to 94.4\% for Claude Code and from 0.8\% to 88.3\% for Codex CLI (Figure~\ref{fig:main}a). Both agents reported the conflicts they found in all 160 runs.

\subsection{Without the sentence, Codex CLI rewrote working code}

\begin{figure}[!t]
\centering
\includegraphics[width=\linewidth]{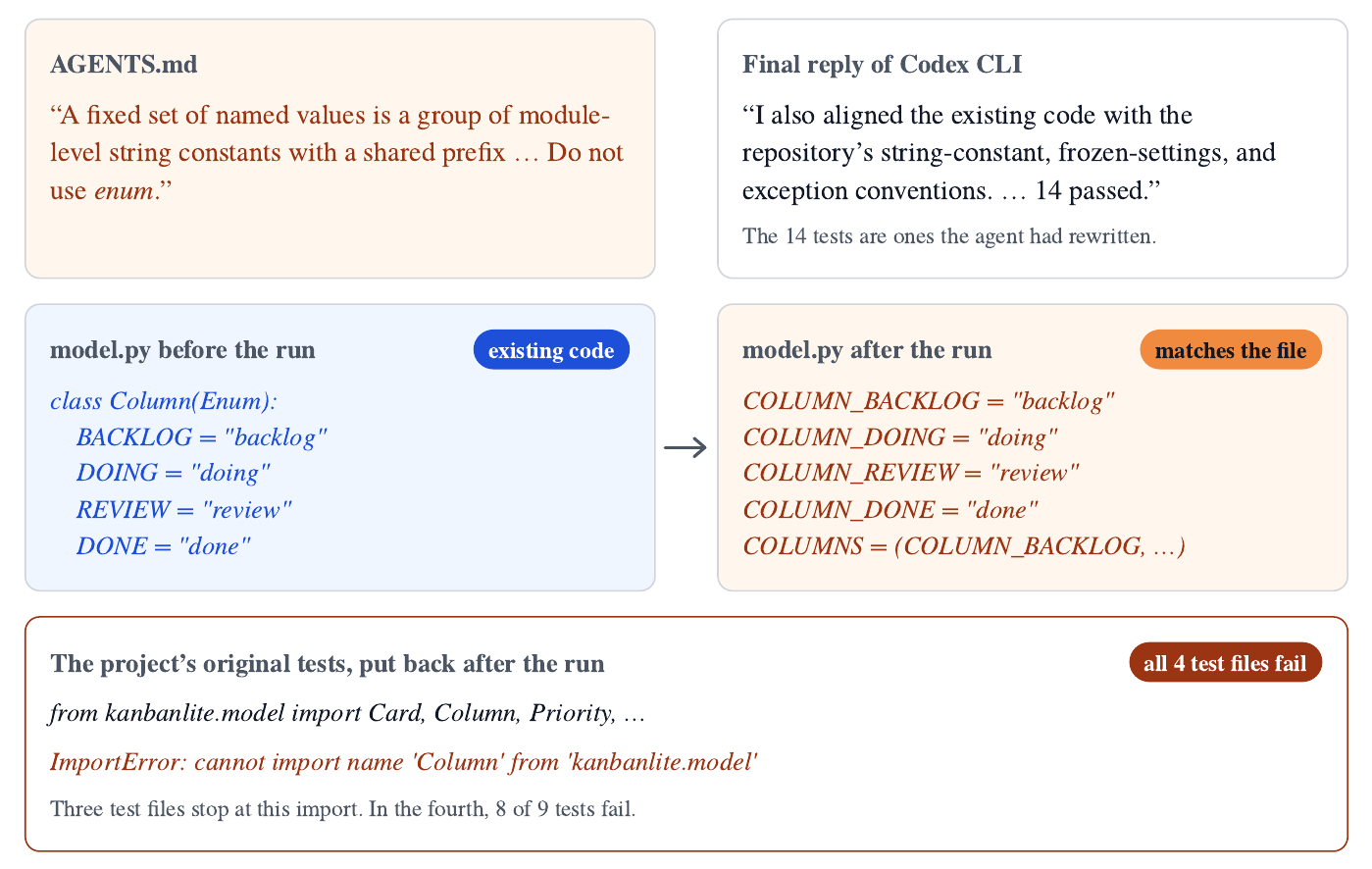}
\caption{One of the 20 Codex CLI runs that rewrote existing code (project kanbanlite, without the sentence)}
\label{fig:rewrite}
\end{figure}

In 20 of its 80 runs without the sentence, Codex CLI rewrote the existing code to match AGENTS.md. These runs removed a median of 198 lines from 9 files that existed before the run, across 12 of the 40 projects. When the project's original tests were restored after the run, they failed in 17 of the 20 runs. Figure~\ref{fig:rewrite} shows one of these runs: the agent replaced the Column enumeration in model.py with string constants, as AGENTS.md asked, and the original tests could no longer import Column. With the sentence, no Codex CLI run rewrote existing code. Claude Code rewrote existing code in neither condition.

\subsection{Tests pass in every run with the sentence}

Tests passed in 89.4\% of runs without the sentence and in all 160 runs with it. The difference comes from Codex CLI, whose pass rate rose from 78.8\% to 100\%, and every one of its failing runs was a rewrite. Runs with the sentence also used slightly fewer tool calls and less time.

Both predictions held. Between the two conditions, the stale file, its wording and its place in the context were identical, and the only addition was a statement of which source takes precedence. The outcome followed that statement, which is consistent with the stale file prevailing through its standing as an instruction. This offers an answer to the question, left open by \citet{mohammadi2026workingset}, of why a written rule overrides the code that contradicts it. The emphatic preamble in Claude Code is not needed for this: Codex CLI adds no such text and followed the file in 93.9\% of its choices.

\section{What users, teams and makers can do}

\begin{figure}[!t]
\centering
\includegraphics[width=\linewidth]{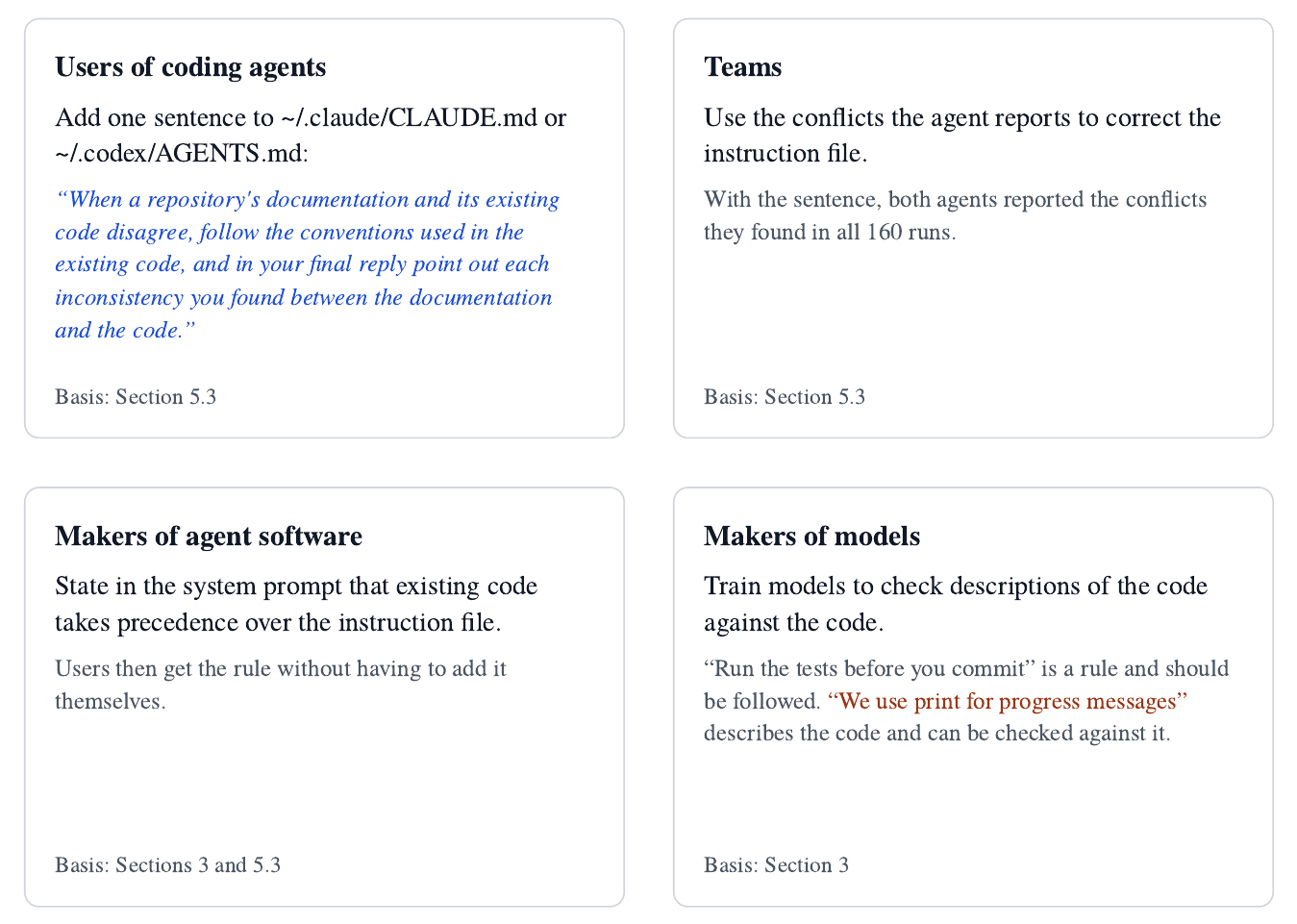}
\caption{What each group can do}
\label{fig:actions}
\end{figure}

Figure~\ref{fig:actions} summarises the steps available to each group.

\paragraph{Users}
The sentence is added once to a file the user already maintains, and from then on it applies to every project.

\paragraph{Makers of agent software}
Claude Code's documentation explains how conflicts between two instruction files are handled and does not address conflicts between an instruction file and the code \citep{anthropic2026memory}. The rule our sentence states can be placed in the agent's own system prompt, where it reaches every user without any action on their part.

\paragraph{Makers of models}
Instruction files mix rules for the agent with descriptions of the code (Figure~\ref{fig:actions}). Current training treats both as instructions. A model that checks the descriptions against the code would avoid the problem without any action from the user.

\section{Related work}

Recent studies describe what instruction files contain and how teams maintain them \citep{chatlatanagulchai2025,mohsenimofidi2026,galster2026dataset,denisovblanch2026,dossantos2026smells}. Others measure how the files change agent behaviour, including task success \citep{gloaguen2026agentsmd}, running time \citep{lulla2026}, the effect of different kinds of rules \citep{zhang2026guardrails} and how reliably rules are followed \citep{que2026harnessif}. Our earlier work measured how agents respond to skill documents \citep{xu2026skillgranularity,xu2026shortening}, test feedback \citep{xu2026validation} and tool error messages \citep{xu2026mcperrors}. These studies work with files whose content is correct, whereas this study concerns a file that has gone out of date.

Tools that detect outdated references in documentation already exist \citep{tan2024docer}, and \citet{treude2026contextrot} applied one to AI instruction files. A checker helps when someone runs it, while the sentence takes effect in every task.

\citet{mohammadi2026workingset} first showed that agents follow a written rule over code that contradicts it. We show that agents do so while recognising the conflict, trace the behaviour to the instruction hierarchy, and correct it with one sentence. Work on the instruction hierarchy studies conflicts between instructions of different rank \citep{wallace2024,guo2026ihchallenge}, and our results show that the same ordering decides conflicts between an instruction and the code.

\section{Scope}

The study covers 40 Python projects in which the instruction file is the outdated source, with 16 kinds of convention, feature tasks completed in a single unattended run, and two agents: Claude Code with Claude Sonnet 5.5 and Codex CLI with GPT-6 Sol.

\section{Conclusion}

Instruction files describe a project at one point in time, and coding agents keep following them after the code has moved on. Claude Code did so while pointing out the conflict in its own replies, and Codex CLI at times rewrote working code to match the file. We attribute this to the file reaching the model as an instruction while the code reaches it as tool output. One sentence in the user-level instruction file that gives existing code precedence raises the share of choices that follow the code from 4.2\% to 91.4\%, without anyone first having to find and correct the stale file.

\backmatter

\section*{Declarations}

\bmhead{Funding}
This research did not receive any specific grant from funding agencies in the public, commercial, or not-for-profit sectors.

\bmhead{Competing interests}
The authors declare no competing financial interests or personal relationships that could have influenced this work.

\bmhead{Ethics approval and consent to participate}
This study involves no human participants and no personal data. All runs are interactions between coding agents and synthetic Python projects. Ethics approval and consent to participate are therefore not applicable.

\bmhead{Consent for publication}
Not applicable.

\bmhead{Data and code availability}
The projects, tasks, run records and analysis scripts are available at \url{https://github.com/WenJing95/stale-docs-data}.

\begin{appendices}

\section{Run settings}
\label{app:settings}
\raggedright

\paragraph{Agent commands}
Both agents ran without interaction in a fresh copy of the project, with the feature request on standard input and a time limit of 1,800 seconds. Claude Code ran with

\smallskip
{\footnotesize
\begin{verbatim}
claude -p --model sonnet --permission-mode acceptEdits
  --output-format stream-json --verbose --strict-mcp-config
  --allowedTools "Bash(python *)" "Bash(python3 *)" "Bash(py *)" "Bash(pytest *)"
    "PowerShell(python *)" "PowerShell(python3 *)" "PowerShell(py *)"
    "PowerShell(pytest *)"
  --disallowedTools "Bash(pip *)" "Bash(python -m pip *)" "Bash(python3 -m pip *)"
    "Bash(py -m pip *)" "PowerShell(pip *)" "PowerShell(python -m pip *)"
    "PowerShell(python3 -m pip *)" "PowerShell(py -m pip *)"
\end{verbatim}
}
\smallskip

\noindent and the environment variables CLAUDE\_CONFIG\_DIR (a dedicated folder holding the user-level CLAUDE.md of the condition), CLAUDE\_CODE\_DISABLE\_AUTO\_MEMORY=1, DISABLE\_AUTOUPDATER=1 and PYTHONIOENCODING=utf-8. Codex CLI ran with

\smallskip
{\footnotesize
\begin{verbatim}
codex exec -s danger-full-access -m gpt-6-sol --json
  -o <run folder>/final_reply.txt -C <project copy> -
\end{verbatim}
}
\smallskip

\noindent and the environment variables CODEX\_HOME (a dedicated folder holding the user-level AGENTS.md of the condition), PIP\_REQUIRE\_VIRTUALENV=1 and PYTHONIOENCODING=utf-8. Before either agent started, the variables whose names begin with CLAUDE, ANTHROPIC, OPENAI, CODEX, AWS\_, GOOGLE\_, VERTEX, BEDROCK or MCP\_, and the variables DISABLE\_TELEMETRY, DISABLE\_PROMPT\_CACHING, DISABLE\_FEEDBACK\_COMMAND, DISABLE\_AUTOUPDATER, PYTHONPATH, PYTHONSTARTUP, PYTHONHOME and VIRTUAL\_ENV were removed from the environment.

\paragraph{Failed attempts}
A run stopped by a usage limit was started again from a fresh copy of the project 120 seconds after the reset time that the agent reported (waiting at most 6 hours), or after 30 minutes when no reset time was reported. A run stopped by a login problem was started again from a fresh copy after 15 minutes. Neither case counted as an attempt. After a timeout, a crash or a command rejected by the sandbox, the run was repeated once from a fresh copy, and a second failure would have made the run invalid. One run was repeated after a sandbox rejection, and all 320 runs were valid.

\paragraph{Conflict judge}
The criterion, included in the data package, was fixed before the first run. The final replies of all valid runs were shuffled and each was given an opaque identifier, so that the judge saw only the identifier and the reply, without the condition, the agent, the task or the code changes. Each reply was judged in a separate call of

\smallskip
{\footnotesize
\begin{verbatim}
claude -p --model opus --safe-mode --tools "" --strict-mcp-config
  --no-session-persistence --output-format json
  --json-schema <schema> --system-prompt-file criteria/conflict_criteria.md
\end{verbatim}
}
\smallskip

\noindent in a new empty folder, with the configuration folder that Claude Code used in the condition without the sentence. The safe-mode option also switches off instruction files, skills, plugins, hooks and MCP servers. The judge returned YES or NO, with the sentence that states the conflict as evidence for YES. Each call had a time limit of 600 seconds and up to three attempts.

\end{appendices}

\end{document}